\documentclass[aps,pra,twocolumn,longbibliography]{revtex4-2}

\usepackage{graphicx}
\usepackage{braket}
\usepackage{amsmath}
\usepackage{amssymb}
\usepackage{amsthm}
\begin{document}
\title{Security of entanglement-based QKD\\with realistic parametric down-conversion sources}
\author{K. S. Kravtsov$^{1,2}$}\email{kk@quantum.msu.ru}
\affiliation{
$^1$ MSU Quantum Technology Centre, Moscow, Russia\\
$^2$ HSE University, Moscow, Russia
}

\date{\today}
\begin{abstract}
The paper analyzes security aspects of practical entanglement-based quantum key distribution (QKD), namely, BBM92 or entanglement-based BB84 protocol.
Similar to prepare-and-measure QKD protocols, practical implementations of the entanglement-based QKD have to rely upon non-ideal photon sources.
A typical solution for entanglement generation is the spontaneous parametric down-conversion. However, this process creates not only single photon pairs,
but also quantum states with
more than two photons, which potentially may lead to security deterioration. We show that this effect does not impair the security of entanglement-based QKD systems.
We also review the available security proofs and show that properties of the entanglement source have nothing to do with security degradation.
%We provide practical examples showing that the presence of higher-order components have nothing to do with security degradation.
\end{abstract}
%\pacs{03.67.Bg, 03.67.Mn, 42.65.Lm}
\maketitle

\section{Introduction}
Quantum cryptography is a mature research direction that resulted in a great number of practical solutions for highly secure cryptographic key distribution.
Security of QKD protocols relies upon the fundamental laws of quantum mechanics. Therefore, their proper practical implementation may achieve the highest degree of key security,
provided all the conventional requirements for cryptographic equipment are met. However, to achieve this level of security, all possible deviations between a particular implementation and the underlying
theoretical model must be scrutinized to make sure they do not open backdoors for an eavesdropper Eve.

Most practical QKD systems are now based on prepare-and-measure protocols, often, on some modifications of the original BB84~\cite{BB84} protocol. One of the weakest spots
in such implementations is unavoidable substitution of true single photons with weak coherent pulses, which may contain more than one photon. Left untreated, this substitution undermines
the security of the original protocol and makes such implementation pointless. Fortunately, effective counter-measures, such as using so-called decoy states, have been found and
are relied upon virtually in all developed systems.

An alternative approach to QKD is to use entangled quantum systems, e.g. EPR photon pairs, to generate cryptographic keys between two agents, Alice and Bob, who perform measurements
of the entangled components. However, a similar conceptual problem becomes a significant obstacle: practically generated entanglement may differ from the ideal EPR pairs. Namely,
instead of generating photon pairs one-at-a-time, more then one pair may be generated, which could potentially lead to an information leak, rendering the system insecure.
Our goal in this paper is to analyze this discrepancy between the basic model and the practical implementation.

We analyze the most straightforward entanglement-based protocol developed in~\cite{BBM92}, which we will call BBM92. Sometimes it is also called an entanglement-based version
of BB84~\cite{SBC09}.
We also assume that entangled states of photons are generated by means of spontaneous parametric down-conversion (SPDC) process, which is by far the most common way
of entanglement generation in the field of quantum optics and quantum cryptography. In particular, we are interested in how multi-photon states generated in SPDC affect the
security of practical BBM92 implementation.

Apparently, there is a vast literature on the entanglement-based QKD, which also gives the correct answer to the question raised above. However, some works~\cite{DB02,BLP07},
which are supposed to eliminate the ambiguity in this topic, do not fully serve this purpose.
Moreover, there is a significant gap between the general security proof, which provides a clear answer, but is non-constructive,
and the low-level understanding of the way the entanglement works.
In this paper we choose the bottom-to-top approach to fill this gap and provide meaningful results for understanding entanglement-based QKD. We pursuit the goal of
presenting a self-contained low-level analysis that is merged with high-level concepts used in the available security proofs.

The paper is organized as follows. In section II we derive expressions for the entangled quantum states generated by SPDC. In section III we analyze the generated four-photon
quantum states. We provide expressions for their representations in both bases used in the QKD protocol. Section IV is devoted to the security analysis and discussion
of the obtained results. We conclude in section V.

\section{Parametric down-conversion}\label{sec_spdc_statistics}
Parametric down-conversion is a nonlinear process where a high energy photon gets effectively split into two photons with lower energy. As the energy is conserved, the sum of the energies
of generated photons equals that of the original (pump) photon. We will focus on the frequency-degenerate SPDC, where the energy of the pump photon is split into two equal parts.
Another assumption is that the two created photons belong to different spatial modes. This is necessary for transmission of the photons to different places --- to Alice and Bob.

To describe this process we will use the second-quantized representation and will call $a^\dag$ and $a$ --- the creation and annihilation operators acting in the spatial mode going to Alice
and $b^\dag$ and $b$ --- the same for the mode going to Bob.
The desired entanglement is typically obtained in the degree of freedom describing the photon polarization. As it is a two-dimensional complex Hilbert space, each spatial mode contains two principal polarizations, which we will call $H$ and $V$ assuming horizontal and vertical polarizations respectively. Thus, creation and annihilation operators in each spatial mode
have an additional subscript defining a particular polarization mode.

The target quantum state needed for BBM92 QKD protocol is $(\ket{HV}-\ket{VH})/\sqrt{2} = (a^\dag_H b^\dag_V - a^\dag_Vb^\dag_H)\ket{0}/\sqrt{2}$,
where $\ket{0}$ is the vacuum quantum state.
This particular Bell state, the singlet state, is chosen here because it is invariant under the change of the polarization basis. However, this state can be transformed
into any other Bell state or some certain linear combinations of them by local polarization transformations in one of the spatial modes. Therefore, this particular choice
can be made without loss of generality.

The corresponding SPDC is described by an action of the Hamiltonian~\cite{KB00}
\begin{equation}\label{eq_hamilton}
	\hat{H} =i\hbar g(a^\dag_H b^\dag_V - a^\dag_Vb^\dag_H) - i\hbar g^*(a_H b_V - a_V b_H)
\end{equation}
upon vacuum state, where $g$ is a complex amplitude of the interaction that depends on the nonlinearity, the pump power, and the spatial configuration.

Using Heisenberg representation it is straightforward to find the evolution of operators in time:
\begin{align}
	\frac {d a_H}{dt} &= \frac i\hbar \bigl[\hat{H}, a_H\bigr] = gb^\dag_V,\\
	\frac {d b^\dag_V}{dt} &= \frac i\hbar \bigl[\hat{H}, b^\dag_V\bigr] =  g^* a_H,\\
	\frac {d a_V}{dt} &= \frac i\hbar \bigl[\hat{H}, a_V\bigr] = -gb^\dag_H,\\
	\frac {d b^\dag_H}{dt} &= \frac i\hbar \bigl[\hat{H}, b^\dag_H\bigr] =  -g^* a_V.
\end{align}
Pairs $b_V$ -- $a^\dag_H$ and $b_H$ -- $a^\dag_V$ behave the same way as $a_H$ -- $b^\dag_V$ and $a_V$ -- $b^\dag_H$, respectively.

Assuming $g = |g|e^{i\varphi}$, get the following solutions
\begin{align}
	a_H(t) &= a_H(0)\cosh|g|t + e^{i\varphi} b^\dag_V(0)\sinh|g|t,\label{eq_atrans1}\\
	b^\dag_V(t) &= e^{-i\varphi} a_H(0)\sinh|g|t + b^\dag_V(0)\cosh|g|t,\\
	a_V(t) &= a_V(0)\cosh|g|t - e^{i\varphi} b^\dag_H(0)\sinh|g|t,\label{eq_atrans3}\\
	b^\dag_H(t) &= -e^{-i\varphi} a_V(0)\sinh|g|t + b^\dag_H(0)\cosh|g|t.\label{eq_atrans4}
\end{align}

Now take a look at the evolution of the quantum state in Schr\"odinger representation
\begin{equation}
	\ket{\Psi(t)} = \exp \left(-\frac{i\hat{H}t}\hbar\right)\ket{\Psi(0)} = \hat{S}(gt)\ket{\Psi(0)}.
\end{equation}
The system in question starts from the vacuum state and experiences the action of $\hat{H}$ for a certain time $t$. Denote $\xi = gt$, and $\ket{\xi} = \hat{S}(\xi)\ket{0}$.
Have
\begin{equation*}
	0 = a_H \ket{0} = \hat{S}(\xi) a_H \hat{S}^\dag(\xi)\hat{S}(\xi)\ket{0} = \hat{S}(\xi) a_H \hat{S}^\dag(\xi)\ket{\xi},
\end{equation*}
and the same for the other three annihilation operators.
Using (\ref{eq_atrans1}) and (\ref{eq_atrans3}) for the transformed annihilation operators $\hat{S}(\xi) a_H \hat{S}(\xi)$, obtain
\begin{align}
	\bigl( a_H\cosh|\xi| + e^{i\varphi} b^\dag_V \sinh|\xi|\bigr)\ket{\xi} &= 0,\label{eq_recurs1}\\
	\bigl( a_V\cosh|\xi| - e^{i\varphi} b^\dag_H \sinh|\xi|\bigr)\ket{\xi} &= 0,\label{eq_recurs2}\\
	\bigl( b_H\cosh|\xi| - e^{i\varphi} a^\dag_V \sinh|\xi|\bigr)\ket{\xi} &= 0,\label{eq_recurs3}\\
	\bigl( b_V\cosh|\xi| + e^{i\varphi} a^\dag_H \sinh|\xi|\bigr)\ket{\xi} &= 0.\label{eq_recurs4}
\end{align}

These equations help to find the final quantum state.
We will use the occupation number representation for the four modes in question. Denote $\ket{ijkl}\equiv \ket{i}_{AH} \ket{j}_{AV} \ket{k}_{BH} \ket{l}_{BV}$,
$\ket{\xi} \equiv \sum_{i,j,k,l} C_{ijkl} \ket{ijkl}$.
From the form of the interaction Hamiltonian it is clear, that photons get created only in pairs. Thus, non-zero contributions appear only from terms where $i = l$ and $j = k$.

Looking at coefficients by the same number states in (\ref{eq_recurs1}) and (\ref{eq_recurs2}) we obtain the following recursive expressions
\begin{align*}
	\sqrt{i+1}C_{i+1\,jkl}\cosh|\xi| + e^{i\varphi}\sqrt{l}C_{ijk\,l-1} \sinh|\xi| &= 0,\\
	\sqrt{j+1}C_{i\,j+1\,kl}\cosh|\xi| - e^{i\varphi}\sqrt{k}C_{ij\,k-1\,l} \sinh|\xi| &= 0.
\end{align*}
They are equivalent to
\begin{align}
	C_{i\,j+1\,k+1\,l} &= e^{i\varphi} \tanh|\xi| C_{ijkl},\\
	C_{i+1\,jk\,l+1} &= - e^{i\varphi} \tanh|\xi| C_{ijkl},
\end{align}
which is enough for writing down the final expression for $\ket{\xi}$. Denote $C_0 \equiv C_{0000}$, obtain
\begin{multline}
	\ket{\xi} = C_0\Bigl[\ket{0000} + e^{i\varphi} \tanh|\xi| \Bigl(\ket{0110} - \ket{1001}\Bigr) +\\
	e^{2i\varphi} \tanh^2|\xi|\Bigl( \ket{0220} - \ket{1111} + \ket{2002}\Bigr) +\\
	e^{3i\varphi} \tanh^3|\xi|\Bigl( \ket{0330} - \ket{1221} + \ket{2112} - \ket{3003}\Bigr) + \dots \Bigr].\label{eq_finalxi}
\end{multline}

The final step is finding $C_0$, which provides a proper norm of the quantum state (\ref{eq_finalxi})
\begin{equation}
	C_0^2\sum_{k=0}^\infty (k+1)(\tanh^2|\xi|)^k = C_0^2\cosh^4|\xi| = 1,
\end{equation}
so $C_0 = 1/\cosh^2|\xi|$.

We arrived at the essentially the same expression that was used in the model~\cite{MFL07}.

\section{Four-photon component}
Before proceeding to the general security considerations let us take a closer look at the four-photon component of $\ket{\xi}$. It is the first and the most
significant term that goes beyond a single photon pair, so its possible effect on the security of QKD is the most prominent.

This component corresponds to the quantum state of
\begin{equation}
	\ket{\psi_4} = \frac 1{\sqrt{3}} \Bigl( \ket{0220}+\ket{2002} - \ket{1111}\Bigr).
\end{equation}
It may seem that while the component $\ket{1111}$ is somewhat neutral, the other two components directly threaten the security of the system, as Eve may get access to the
second pair of photons with the same polarizations as for Alice and Bob.
However, the ingenuity of BBM92/BB84 protocols is in measurements in randomly chosen mutually unbiased bases.
If the measurement basis was known to Eve, she could separate these three components by performing non-demolition measurement of the photon numbers in different polarization modes,
blocking the unfavorable $\ket{1111}$ states. In reality, this will result in delivering polarized states to Alice and Bob, and therefore, uncorrelated measurement results,
should they choose the opposite basis.

For the better demonstration we will find representations of these
quantum states in the diagonal basis $\ket{D}$, $\ket{A}$. The corresponding creation operators are defined by
\begin{align}
	a^\dag_D &= \frac1{\sqrt{2}} \bigl( a^\dag_H + a^\dag_V \bigr),\\
	a^\dag_A &= \frac1{\sqrt{2}} \bigl( a^\dag_H - a^\dag_V \bigr),
\end{align}
and the same for the operators on the Bob's side.

For simplicity we will use a notation where primed quantum states are written in D-A basis:
$\ket{ijkl}'\equiv \ket{i}_{AD} \ket{j}_{AA} \ket{k}_{BD} \ket{l}_{BA}$, where the first subscript is the spatial mode and the second --- the polarization one.
As Alice's and Bob's sides are independent of each other, let us first find D-A representations for Alice's subsystem:
\begin{align*}
	\ket{11} &= a^\dag_H a^\dag_V\ket{0}  = \frac {{a^\dag_D}^2 - {a^\dag_A}^2} 2\ket{0} = \frac{\ket{20}' - \ket{02}'}{\sqrt{2}},\\
	\ket{20} &= \frac{{a^\dag_H}^2}{\sqrt{2}}\ket{0} = \frac{\bigl( a^\dag_D + a^\dag_A\bigr)^2}{2\sqrt{2}}\ket{0}  =  \frac{\ket{20}'}{2} + \frac{\ket{11}'}{\sqrt{2}} + \frac{\ket{02}'}{2},\\
	\ket{02} &= \frac{{a^\dag_V}^2}{\sqrt{2}}\ket{0} = \frac{\bigl( a^\dag_D - a^\dag_A\bigr)^2}{2\sqrt{2}}\ket{0}  =  \frac{\ket{20}'}{2} - \frac{\ket{11}'}{\sqrt{2}} + \frac{\ket{02}'}{2}.\\
\end{align*}
The first expression in fact illustrates Hong–Ou–Mandel effect in the polarization domain.

Now, expressions for the whole 4-photon component are
\begin{multline}
	\frac{\ket{0220} + \ket{2002}}{\sqrt{2}} = \frac 1{2\sqrt{2}}\Bigl( \ket{2020}' + \ket{2002}' \\
	+ \ket{0220}' + \ket{0202}' - 2\ket{1111}'\Bigr),
\end{multline}
\begin{equation}
	\ket{1111} = \frac 12\Bigl(\ket{2020}' - \ket{2002}' - \ket{0220}' +\ket{0202}'\Bigr).
\end{equation}

Finally, we can easily see that
\begin{equation}
	\ket{\psi_4} = \frac 1{\sqrt{3}} \Bigl( \ket{0220}'+\ket{2002}' - \ket{1111}'\Bigr).
\end{equation}
That is, the 4-photon SPDC component is invariant to the basis change from H-V to D-A. In this sense it behaves the same way as the 2-photon singlet state.
It appears that the interaction Hamiltonian~(\ref{eq_hamilton}) itself is invariant to such a basis change, so any higher-order component must be invariant as well.

The next step towards the discussion about the protocol security is understanding how Eve can attack these 4-photon states.
Let us take a look at the quantum channel from the source to Alice. If Eve performs an intercept-and-resend attack, she will not be able to deliver correct key bits if
she happens to use a wrong basis. This situation is the same as for the conventional BB84 protocol, where this strategy obviously fails.

The only viable option is to split the two-photon state in this quantum channel retaining one of the photons. It may be stored until bases are disclosed and then
it can be measured in the correct basis.
Another photon must remain in the channel.
So, Eve should create one more spatial mode and split the original state such as one photon remains in the quantum channel and the other one is places into the new mode.
It can be done in the following way:
\begin{enumerate}
	\item Eve places a symmetric beam splitter into the channel and performs a non-demolition measurement of the number of deflected photons;
	\item if it is exactly one, she has obtained the desired result;
	\item if the result is zero or two, she places the component with two photons back into the original mode and starts over from item 1.
\end{enumerate}
As non-demolition measurements of photon numbers act on the whole spatial mode, they do not disturb the polarization degree of freedom, so the polarization states remain intact.

The beamsplitter she uses is described by the substitution
\begin{equation}
	a^\dag_{AX} \rightarrow \frac 1{\sqrt{2}}\bigl(a^\dag_{AX} + a^\dag_{EX}\bigr),
\end{equation}
where $X$ is either $H$ or $V$, and $E$-mode is the one captured by Eve.

A polarization mode with two photons trivially splits into the two spatial modes with the same polarization. The success probability for Eve is 1/2.

A spatial mode with two orthogonally polarized photons is a bit more complicated:
\begin{multline}
	a^\dag_{AH}a^\dag_{AV}\ket{0} \rightarrow \frac 12 \Bigl(a^\dag_{AH}a^\dag_{AV} + a^\dag_{EH}a^\dag_{EV}\\
	+ a^\dag_{EH}a^\dag_{AV} + a^\dag_{AH}a^\dag_{EV}\Bigr) \ket{0}.
\end{multline}
The terms on the first line correspond to a failed attempt of separating the photons, while the last two terms describe the state after successful separation, which
again happens with the probability of 1/2.

Now we switch from the occupation number representation to the polarization representation as each spatial mode now contains exactly one photon.
So the successfully split state in this notation becomes $(\ket{V}_A\ket{H}_E + \ket{H}_A\ket{V}_E)/\sqrt{2}$.

The Bob's quantum channel may be treated by Eve the same way, and the resulting overall quantum state becomes
\begin{multline}
	\ket{\psi_4} \rightarrow \frac 1{2\sqrt{3}}\biggl[ \ket{HV}_{AB}\Bigl(2\ket{HV}_E - \ket{VH}_E\Bigr)\\
	+ \ket{VH}_{AB}\Bigl(2\ket{VH}_E - \ket{HV}_E\Bigr)
	-\ket{HH}_{AB}\ket{VV}_E\\
	- \ket{VV}_{AB}\ket{HH}_E \biggr].\label{eq_fpolstate}
\end{multline}
The same expression may be written for this quantum state in the diagonal basis and it will have exactly the same form due to the aforementioned symmetry.

We can now easily analyze this result by calculating QBER for the channels Alice-Bob and Alice-Eve. For simplicity we assume that all three characters possess
ideal single-photon detectors. With the desired singlet Bell pairs and without Eve, Alice and Bob always get perfectly anti-correlated measurement results.
The state (\ref{eq_fpolstate}) results in the QBER$_{AB}$ of $1/6$.

We can also consider the case where Eve performs her measurement in the diagonal basis, while Alice and Bob use the basis H-V. In this case it is straightforward to show
that Eve will have zero correlation with Alice's results, so the measurement in the wrong basis is absolutely meaningless.

\section{Protocol security}

Following the chosen bottom-to-top approach we start this discussion from analyzing security implications of the obtained results and then turn to the general picture.

The elegance of QKD is in a mutual relation between the knowledge Eve can obtain from her measurements and the observed quantum bit-error ratio (QBER). Zero QBER indicates
no information leak to Eve, while a non-zero QBER must be associated with such a leak. The leaked information must be taken care of at the privacy amplification step. The raw key should be
shrunk via hashing to ensure arbitrarily small dependence on data available to Eve.
In the analysis below we will stick to the asymptotic case of arbitrarily long measured sequences, which gives the correct picture without unnecessary technical difficulties.

For both BB84 and BBM92 protocols the observed QBER of $Q$ implies that $h(Q)$ per channel use is the upper bound on the amount of leaked information~\cite{GLLP04},
where $h(x) = -x\log_2 x - (1-x)\log_2 (1-x)$ is the binary entropy function.

Given the joint quantum state (\ref{eq_fpolstate}) Eve's best strategy is to wait until bases are disclosed and then analyze her qubits.
In the case Alice and Bob in the result of a measurement obtain the same polarizations, which happens with probability of 1/6, Eve receives quantum states
$\ket{HH}_E$ or $\ket{VV}_E$, which are orthogonal to her other outcomes.  That is, she clearly sees locations of errors in data sequences of Alice and Bob.
That, however, does not tell her any information about the
key bits that Alice and Bob generate when their measurement results differ.

In this case, in order to find the bits of the key, Eve should distinguish between two non-orthogonal quantum states of $\bigl(2\ket{HV}_E - \ket{VH}_E\bigr)/\sqrt{5}$
and $\bigl(2\ket{VH}_E - \ket{HV}_E\bigr)/\sqrt{5}$.
Obtaining information from a random sequence of such states is equivalent to a binary classical-quantum channel, whose capacity equals the corresponding Holevo quantity~\cite{H73,H98}.
In the case of a binary channel Holevo quantity equals
\begin{equation}
	\chi = h\left(\frac {1-|c|}2\right),
\end{equation}
where $c$ is a scalar product between the two quantum states used in the channel. In our case we obtain $\chi = h(1/10)$, while the upper bound on the information leak
accounted for by Alice and Bob is $h(Q) = h(1/6)$.

We found that quantum state (\ref{eq_fpolstate}) is sub-optimal for Eve, as her available information is strictly smaller than it could be at the same level of
the QBER observed by the legitimate interlocutors.
The same holds for a more typical case where such 4-photon states appear in the channel with a probability $p$. The observed QBER is then $p/6$, while Eve's information
is bounded by $p\,h(1/10)$. For any $p$ the information available to Eve is strictly smaller than the boundary Alice and Bob account for. Therefore, there is clearly no security
threat associated with the presence of the 4-photon component in the SPDC signal.

Unfortunately, there exist enough controversion on this topic in the literature. Publications~\cite{DB02,BLP07,BHB09} and even a review~\cite{SBC09} suggest that there is a
significant difference between the coincident independently generated pairs and genuine 4-photon states; and the latter may lead to security impairments.
In reality, neither can cause any risks to the system security. At the same time, the mentioned difference between multi-pair events of the two types is real~\cite{SRM05}
and was experimentally measured in~\cite{THT04,O05}.

Luckily, there are publications~\cite{WZY02,ART08,SBC09,LCT14,XMZ20} where a more correct point of view is present. BBM92 protocol is immune against photon number splitting and even
allows the source of entanglement to be under full Eve's control.
Actually, if it wasn't compatible with untrusted entanglement sources, it would only be usable in the scenario, where the source is inside Alice's (Bob's) setup.
Indeed, if the source is in an external location even with the trusted environment, Eve is supposed to have access to both quantum channels. She may cut them and replace
with her own source of the signal. In this sense any external source is inevitably under total Eve's control.
Therefore, there is not much practical interest in discussing particular source deviations from the ideal model. The problem is much more general: whether it is possible to construct
quantum states that will leak information to Eve without revealing that fact via the increase of the QBER for legitimate users.

The anticipated strong correlations of measurement results between Alice and Bob in both mutually unbiased bases cannot exist without quantum entanglement. They contradict any local
hidden-variable theory and are closely connected with the violation of Bell inequality. Thus, the grounds for providing security lie in the plane of quantum entanglement properties.
The particular property crucial for the protocol operation is called monogamy of entanglement~\cite{T04,HHH09}. It is the property that limits entanglement of an entangled pair with
the outside world.
Indeed, the maximally entangled bipartite state is a pure state. If it was entangled with something else, it would become mixed due to tracing out the rest of the overall system.

Assume we have a system of qubits: A, B, C, etc.
If we use the concurrence squared as the amount of available entanglement, we can calculate this figure for pairs A-B, A-C, and so on.
It has been proved~\cite{CKW00,KW04,OV06} that their sum cannot exceed the concurrence squared between the qubit A and the rest of the system.
It is easy to see why this is relevant to the operation of the protocol. In order to achieve the error-free operation, qubits of Alice and Bob have
to be in the maximally entangled state with the corresponding concurrence of 1. At the same time, concurrence between the qubit of Alice and the rest of a more complex
quantum system cannot exceed 1. Therefore, there is basically no room for entanglement between Alice's qubit and something else.

Clearly, the even more interesting case is a trade-off between the amount of entanglement between Alice and Bob and the observed QBER. To the best of our knowledge there is no yet known
connection between the two. However, the available security proofs, in particular~\cite{KP03,GLLP04} do assume an arbitrary and uncharacterized entanglement source. The only strict
requirement is that it has to be independent on the measurement basis chosen by Alice and Bob.

Another limitation playing a crucial role in the security of the protocol is that Alice and Bob must perform measurements of single qubits, and not of larger quantum systems.
It can be easily shown~\cite{PAB09}, that a violation of this requirement (e.g. by providing malicious measurement hardware that accepts different qubits depending on the measurement
``basis'') may be used for rendering the system insecure.

In practice, the incoming quantum states are measured by sending them to a polarization beam splitter with a subsequent threshold detection. It is performed by conventional single-photon
detectors not capable of resolving photon numbers. It has been shown in~\cite{BML08} that such detection effectively performs the required ``squashing'' to a single qubit subspace.
In~\cite{TT08} an explicit model for BBM92 protocol is provided. One feature of the model is that the receiver should assign a random value to the detection event when two
detectors fired coincidently, rather than just discarding it. It might be an important point in the security analysis. Such events are rather
rare in most practical settings, however it might make sense to follow this rule.  To the best of our knowledge it is unknown whether there exist an attack exploiting this subtle
feature.

There is an interesting implication of the obtained results. It is well-known that sub-Poissonian signal sources are favorable for realization of prepare-and-measure QKD protocols.
It is reported that heralded single-photon sources based on SPDC are used as a signal source for BB84 protocol~\cite{WCX08,WZL09}. However, due to the presence of the multi-pair
components, discussed in section~\ref{sec_spdc_statistics}, one needs to implement decoy-state protocol on the top of such a single-photon source.
At the same time, if the said SPDC generated entangled states, one could implement entanglement-based BBM92 where the source is in the Alice's setup.
This would make decoy states unnecessary.

Interestingly,
in both cases the actual photon statistics in the channel is the same. The difference is only in the state modulation. The former system actively modulates {\it all} photons in the channel, thus,
making photon number splitting a real threat.
The latter system carries photons independent on the measurement Alice performed. There is no information about Alice's basis and her measurement result in the channel. That makes it oblivious to
the presence of a multi-photon component.

\section{Conclusion}

We demonstrated elements of the entanglement-based QKD protocol operation with a realistic SPDC source. The unavoidable presence of the multi-pair component in the source
in fact causes partial information leak to Eve. We analyzed corresponding probabilities, polarization quantum states, and extent of the key exposure in detail.
At the same time, {\it the mentioned multi-pair component is not, and may not be, a vulnerability} of a QKD system:
the genuine side-effect of this information leak is the unavoidable increase of the QBER measured by the legitimate users. Thus, any Eve's knowledge about the key
(if the final key can be extracted at all) is extinguished with the privacy amplification step.
We reviewed the related security proofs and put together a summary of the natural security requirements for the entangled-based QKD.
{\it The entanglement source quality is, in fact, entirely out of the scope of these requirements, as even an unknown source under total adversary's control may be
used for generation of unconditionally secure keys, provided the observed QBER is below a known bound.}

\section*{Acknowledgments}
This work was supported by the contract with JSCo Russian Railways.

\end{document}